\documentclass[aps,amssymb,showpacs,twocolumn]{revtex4}
\usepackage{amssymb}
\usepackage{graphicx}
\usepackage{color}
\usepackage{textcomp}

\begin{document}

\title{Deconfinement Phase Transition with External Magnetic Field in Friedberg-Lee Model }
\author{Shijun Mao}
\affiliation{School of Science, Xian Jiaotong University, Xian 710049, China}

\begin{abstract}
The deconfinement phase transition with external magnetic field is investigated in the Friedberg-Lee model. In the frame of functional renormalization group, we extend the often used potential expansion method for continuous phase transitions to the first-order phase transition in the model. By solving the flow equations we find that, the magnetic field displays a catalysis effect and it becomes more difficult to break through the confinement in hot and dense medium.
\end{abstract}

\date{\today}
\maketitle

The most interesting quantum
chromodynamics (QCD) phase transition at finite temperature and density is the deconfinement from hadron gas to
quark-gluon plasma (QGP). It is
widely accepted that the quark matter may exist in compact stars with high baryon density and the initial stage of high energy nuclear collisions with high temperature. Recently the study on the QCD phase transitions is extended to including magnetic field, since the strongest magnetic field
in nature may be generated in relativistic
heavy ion collisions. The initial magnitude of
the field can reach $eB\sim (1-100)m_\pi^2$ in collisions at
the Relativistic Heavy Ion Collider and the Large Hadron Collider~\cite{b0,b1,b2,b3,b4}, where $e$ is the electron charge and $m_\pi$ the pion mass in vacuum.
When the magnetic field survives in the later formed hot medium, the
interaction between the field and the medium may change
the fundamental QCD topological structure (see~\cite{rb1,rb2,rb3} for reviews).

Whether the magnetic field can induce deconfinement phase transition is still an open question. From the lattice simulations, the dependence of the Polyakov loop on the magnetic
field supports the scenario of inverse magnetic catalysis, which shows a decreasing critical temperature as the field strength grows~\cite{rb3,bpl1,bpl2}. While in the MIT bag model and Polyakov quark meson model there is also the inverse magnetic catalysis~\cite{mit1,mit2,p1}, the magnetic field displays a catalysis effect in the Polyakov NJL model~\cite{p2} and turns to be an inverse catalysis only when a magnetic field dependent coupling constant is introduced~\cite{p3}. As for the phase transition at finite density, there is not yet precise lattice result due to the well-known sign problem.

The Friedberg-Lee model~\cite{flm1},
also referred as a nontopological soliton model~\cite{flm2}, is widely discussed in the study of confined and deconfined quarks~\cite{mh1,lee,birse,li,gs,yang,mh}. In this model, the nonperturbative
dynamics responsible for confinement in QCD is
simulated in terms of a nonlinear coupling of a scalar field $\sigma$.
In vacuum state, the ensemble average of $\sigma$ is large
and the quark mass is about 1 GeV, the heavy quarks are confined in hadron bags. With
increasing temperature and/or density of the system, the
average of $\sigma$ and in turn the effective quark mass
drops down, the thermodynamic motion leads to the deconfinement
of the effective light quarks.

In this work we extend the Friedberg-Lee model to including an external magnetic field. We apply the functional renormalization
group (FRG) method to the model to study deconfinement phase transition at finite temperature, baryon density and magnetic field. As a nonperturbative method, the FRG~\cite{wett1,wett2} is used to study phase transitions in various systems
like cold atom gas~\cite{cold1,cold2}, nucleon gas~\cite{nucleon}, hadron gas~\cite{hadron1,hadron2,hadron3,hadron4,hadron5,hadron6,hadron7,hadron8}, and quark matter~\cite{quark1,quark2,quark3}. By solving the flow equation which connects
physics at different momentum scales, the FRG shows a
great power to describe the phase transitions and the corresponding
critical phenomena, which are normally difficult
to be controlled in mean-field approximation
because of the absence of quantum fluctuations. Instead
of adding hot loops to the thermodynamic potential in the
usual ways of going beyond mean field, the FRG effective
potential includes
quantum fluctuations through mass, coupling
constant and wave function renormalizations.

The Friedberg-Lee model is defined as~\cite{flm1}
\begin{eqnarray}
\label{fl}
{\cal L} &=& {\bar \psi} \left(i\gamma^\mu \partial_\mu -g \sigma \right)\psi + \frac{1}{2} \partial^\mu \sigma \partial_\mu \sigma -U(\sigma),\nonumber\\
U(\sigma) &=& \frac{a}{2!}\sigma^2 +\frac{b}{3!}\sigma^3 +\frac{c}{4!}\sigma^4 +B
\end{eqnarray}
with $\psi,\ \bar\psi$, and $\sigma$ being quark, antiquark, and scalar fields,
respectively, where we have neglected the current quark mass to simplify the calculations. There are four
parameters in the Friedberg-Lee model, $a$ with dimension
$L^{-2}$, $b$ with dimension $L^{-1}$, dimensionless $c$, and the Yukawa coupling
constant $g$ between quark and scalar fields. The bag constant
$B$ used to provide a quark confinement is defined as the potential difference between perturbative vacuum and physical vacuum, $B=U(\sigma_{\text {per}})-U(\sigma_{\text {phy}})$. The four parameters can be determined by solving the Euler-Lagrange equations and fitting the proton charge radius, proton magnetic moment and the ratio of axial-vector to vector coupling~\cite{flm1,flm2,mh1,gs}. Different parameter sets are listed in Table \ref{para1}~\cite{gs}.

%%%%%%%%%%%%%%%%%%%%%%%%%%%%%%%%%%%%%%%%%%%%%%%%%%%%%%%%%%%%%%%%%%%%%%%%%%%%%%%%
\begin{table}[ht]
\begin{center}
\caption{The Friedberg-Lee model parameters~\cite{gs}.} %[Gao, Wang, Li: PRD 46, 3211 (1992)]}
\begin{tabular}{ccccc}
set \ & $a(\text{fm}^{-2})$\            & $-b(\text{fm}^{-1})$ \           & $c$ \             & $g$  \\
\hline
1\ &  69.73 \ & 2112.6 \ & 20000 \ & 12.416  \\
\hline
2\ &188.6 \ & 7774.0 \ & 100000 \ & 15.333 \\
\hline
3\ & 11.60 \ & 834.4 \ & 10000 \ & 10.957  \\
\hline
4\ &  17.70 \ & 1457.4 \ & 20000 \ & 12.16  \\
\hline
5\ & 45.21 \ & 5208.5 \ & 100000 \ & 16.379 \\
\hline
6\ & 6.734 \ & 778.48 \ & 10000 \ &  10.963  \\
\hline
7\ & 10.25 \ & 1358.4 \ & 20000 \ & 12.211 \\
\hline
8\ & 26.12 \ & 4848.4 \ & 100000 \ & 16.537 \\
\hline
\end{tabular}\\
\label{para1}
\end{center}
\end{table}
%%%%%%%%%%%%%%%%%%%%%%%%%%%%%%%%%%%%%%%%%%%%%%%%%%%%%%%%%%%%%%%%%%%%%%%%%%%%%%%%%%%%

Making a shift $\sigma = \langle\sigma\rangle+\delta\sigma$ for the scalar field, where $\langle\sigma\rangle$ is the ensemble average, and substituting it into the Lagrangian density (\ref{fl}), the
effective potential of the system at classical level is $U(\langle\sigma\rangle)$, and the dynamical quark and sigma masses generated by the mean field $\langle\sigma\rangle$ can
be extracted from the coefficients of the terms $\bar\psi\psi$ and $(\delta\sigma)^2$,
$m_q = g\langle\sigma\rangle$ and $m_\sigma^2 = a+b\langle\sigma\rangle+c/2\langle\sigma\rangle^2$.
The minimization of the classical potential, $\partial U(\langle\sigma\rangle)/\partial\langle\sigma\rangle=0$, which is equivalent to the condition that the term linear in $\delta\sigma$ disappears from the Lagrange density, is called the gap equation of the system and determines the value of
$\langle\sigma\rangle$. From its solution, there are two vacuum states, the perturbative vacuum and physical vacuum, located respectively at $\langle \sigma \rangle|_{\text{per}}=0$
and $\langle \sigma \rangle|_{\text{phy}}=\frac{3|b|}{2c}\left(1+\sqrt{1-\frac{8ac}{3b^2}}\right)$. To simplify the notation, we take $\sigma$ instead of $\langle\sigma\rangle$ in the following.

We now apply the functional renormalization group method to the Friedberg-Lee model. The core quantity in the framework of FRG is the average action $\Gamma_k\left[\sigma,\psi\right]$
The scale $k$ dependence of the average action is characterized by the Wetterich flow equation~\cite{wett1,wett2} in momentum representation,
\begin{eqnarray}
\label{flow}
\partial_k \Gamma_k &=& \frac{1}{2}\text {Tr} \left[\left(\Gamma^{(2,s)}_k +R_k^s\right)^{-1}  \frac{\partial  R_k^s}{\partial k} \right]\nonumber\\
&&-\text {Tr} \left[\left(\Gamma^{(2,q)}_k +R_k^q\right]^{-1}  \frac{\partial  R_k^q}{\partial k} \right],
\end{eqnarray}
where the trace $\text{Tr}$ is defined in the inner and momentum space, $\Gamma^{(2,s)}_k$ and $\Gamma^{(2,q)}_k$ are the functional derivatives of the average action $\Gamma^{(2,s)}_k=\delta^2\Gamma_k/\delta\sigma^2$ and $\Gamma^{(2,q)}_k=\delta \Gamma_k/\delta (\bar\psi\psi)$, and the infrared cutoff functions $R_k^s$ and $R_k^q$ for scalar and quark fields, which are used to suppress quantum fluctuations at low momentum $p<k$, are chosen as the optimized regulators~\cite{litim,regulator1,regulator2,regulator3}
\begin{eqnarray}
\label{rk}
R_k^s &=& \left(k^2-{\bf p}^2\right)\Theta\left(k^2-{\bf p}^2\right), \\
R_k^q &=&\left({\gamma^\nu p_\nu} +i\mu \gamma^0\right) \sqrt{\frac{(p_0+i\mu)^2+k^2}{(p_0+i\mu)^2+{\bf p}^2}-1}\Theta(k^2-{\bf p}^2).\nonumber
\end{eqnarray}

Taking the local potential approximation, the average action reads
\begin{eqnarray}
\label{gamma}
\Gamma_k &=&\int d^4x \Big(Z^q_k\bar\psi\left(i\gamma^\mu\partial_\mu-g_k\sigma\right)\psi+{1\over 2}Z^s_k\partial^\mu\sigma\partial_\mu\sigma\nonumber\\
&& +U_k(\sigma)\Big),
\end{eqnarray}
where $Z^q_k$ and $Z^s_k$ are the wave functional renormalization constants, $g_k$ is the renormalized Yukawa coupling constant, and $U_k(\sigma)$ corresponds to the potential in the
Lagrangian density (\ref{fl}) but with scale dependent parameters $a_k,\ b_k,\ c_k$ and $B_k$.

Assuming space-time independent ensemble averages, and neglecting the Yukawa coupling constant renormalization and wave function
renormalization ($g_k=g,\ Z_k^q=Z_k^s=1$), the average action to the lowest order is
determined by the potential only~\cite{hadron4,hadron6,hadron7,hadron8,ps0,ps2,ps3,g1}, $\Gamma_k=V_4 U_k(\sigma)$, with the four dimensional space-time volume $V_4$. %To simplify the notation, we use $\sigma$ to replace $\langle\sigma\rangle$ in the following.

With the chosen regulators $R_k^s$ and $R_k^q$, after doing the
three-momentum integration over the sigma and quark fields,
the FRG flow equation (\ref{flow}) can be simplified as
\begin{eqnarray}
\partial_k U_k &=& T \sum_m \frac{k^4}{6\pi^2} \frac{1}{\omega_m^2+k^2+m^2_\sigma}\nonumber\\
&&-4 N_c N_f T\sum_n{\frac{k^4}{6 \pi^2} \frac{1}{\omega_n^2+k^2+m_q^2}},
\end{eqnarray}
where $N_c=3$ and $N_f=2$ are color and flavor numbers of quarks, and $\omega_m=2m\pi T$ and $\omega_n=(2n+1)\pi T$ with $m,n=0,1,2,\cdots$ are the boson and fermion Matsubara frequencies in the imaginary time
formalism of finite-temperature field theory. Doing the frequency summation explicitly, the flow equation can be expressed in terms of the boson and fermion distribution functions $n_b(x)=1/\left(e^x-1\right)$ and $n_f(x)=1/\left(e^x+1\right)$,
\begin{eqnarray}
\label{flow2}
\partial_k U_k &=& {k^4\over 6\pi^2}{1\over E_{\sigma k}}\left({1\over 2}+n_b(E_{\sigma k})\right)\\
&&-2 {k^4\over \pi^2}{1\over E_{qk}}\left(1-n_f(E_{qk}+\mu)-n_f(E_{qk}-\mu)\right),\nonumber
\end{eqnarray}
where $E_{\sigma k}$ and $E_{qk}$ are sigma and quark energies $E_{\sigma k}=\sqrt{k^2+m_\sigma^2}$ and $E_{qk}=\sqrt{k^2+m_q^2}$ with $m_\sigma^2=\frac{\partial^2 U_k}{\partial \sigma^2}$ and $m_q=g\sigma$. Given the initial potential $U_\Lambda$ at the ultraviolet momentum $k=\Lambda$, the potential at any $k$ could be obtained by numerically solving the flow equation (\ref{flow2}).

In the Friedberg-Lee model, the quantum and thermal fluctuations come mainly from quarks which carry larger number of inner freedoms and lighter mass in comparison with the scalar meson. In the flow equation (\ref{flow2}), the vacuum term and medium term of quarks have opposite signs, and the cancelation between them leads to the deconfinement phase transition.

As a simple example, we consider the quantum fluctuations in vacuum where there is no thermal fluctuations. In the spirit of the Friedberg-Lee model, the sigma particle is heavy enough ($m_\sigma > 2$ GeV) to guarantee the quark confinement~\cite{flm1,flm2}. In this case, its fluctuations in vacuum can be safely neglected and the flow equation becomes
\begin{equation}
\label{vacuum}
\partial_k U_k = -2 {k^4\over \pi^2}{1\over E_{qk}}.
\end{equation}
Its analytic solution at $k=0$ can be obtained explicitly,
\begin{eqnarray}
U_0(\sigma) &=& U_\Lambda(\sigma)+{1\over 8\pi^2}\Big[2\Lambda E_{q\Lambda}\left(2\Lambda^2-3(g\sigma)^2\right)\nonumber\\
&&-6(g\sigma)^4\ln{g\sigma\over \Lambda+E_{q\Lambda}}\Big]
\end{eqnarray}
with the initial condition $U_\Lambda(\sigma)$.

We now turn on the external magnetic field ${\bf B}$ in the Friedberg-Lee model and choose ${\bf B}=B{\bf e}_z$ without loss of generality. Since $\sigma$ is a neutral scalar field and there is no direct interaction with the magnetic field, we consider only the magnetic effect on the quark part of the flow equation. In the external magnetic field, the quark momentum integration is replaced by an integration over the momentum along the $z-$axis plus a summation over the Landau energy levels on the transverse plane~\cite{landau},
\begin{equation}
\int {d^3{\bf p}\over (2\pi)^3} \to {|Q_fB|\over 4\pi}\sum_i \alpha_i\int{dp_z\over 2\pi}
\end{equation}
with $\alpha_i=2-\delta_{i 0}$, electric changes $Q_u=2e/3$ and $Q_d=-e/3$, and momentum ${\bf p}^2=p_z^2+2|Q_fB|i$. With this replacement, the flow equation (\ref{flow2}) becomes
\begin{eqnarray}
\label{flow3}
\partial_k U_k &=& {k^4\over 6\pi^2}{1\over E_{\sigma k}}\left({1\over 2}+n_b(E_{\sigma k})\right)\\
&&-{3\over 2\pi^2}\sum_{s=\pm 1}\sum_{f=u,d}\sum_{i=0}^\infty{|Q_fB|k\over E_{qk}}\sqrt{k^2-p_\bot^2}\times\nonumber\\
&&\Theta(k^2-p_\bot^2)\left(1-n_f(E_{qk}+\mu)-n_f(E_{qk}-\mu)\right)\nonumber
\end{eqnarray}
with quark transverse momentum $p^2_\bot=(2i+1-s)|Q_f B|$ and the quark energy $E_{qk}=\sqrt{k^2+p^2_\bot+m_{q}^2}$. %Here we have used the assumption that the quark regulator $R_k^q$ is independent of the magnetic field~\cite{rk}.

We solve the flow equation (\ref{flow3}) with the generalized potential expansion method. The potential expansion method is successfully used in the study of chiral symmetry, $U_A(1)$ symmetry and deconfinement at finite temperature~\cite{hadron4,hadron6,hadron7,hadron8,ps0,ps2,ps3}, provided that the phase transition is continuous which guarantees only one minimum of the potential. However, due to the character of the first-order phase transition in the Friedberg-Lee model, we should separately expand the effective potential around the two local minima, and the comparison between them determines the real ground state of the system. This is a generalization of the potential expansion method from second- to first-order phase transitions.

We expand the potential $U_k(\sigma)$ on the left-hand side of Eq.(\ref{flow3}) around the two local minima $U(\sigma_k)$ which satisfy the gap equation $\partial U_k(\sigma_k)/\partial\sigma_k=0$ and $\partial^2 U_k(\sigma_k)/\partial\sigma_k^2 \ge 0$. Shifting the field $\sigma\to\sigma_k+\delta\sigma_k$, $\partial_k U_k$ could be expressed in terms of the powers of $\delta\sigma_k$,
\begin{equation}
\label{lpe}
\partial_k U_k(\sigma)=\sum_{j=0}^4 {1\over j!}{\partial^j \dot U_k(\sigma)\over \partial\sigma^j}\Big|_{\sigma=\sigma_k}(\delta\sigma_k)^j.
\end{equation}
By comparing the coefficients
of $(\delta\sigma_k)^j$ on the
left- and right-hand sides of the flow equation (\ref{flow3}), one
obtains four coupled differential equations for the four
parameters $a_k,\ b_k,\ c_k$ and $B_k$. Note that in deriving the four flow equations,
we have used the relation $\partial_k\delta\sigma_k=-\partial_k\sigma_k$.

Together with
the gap equation, the $k$ dependence of the four parameters and the classical field $\sigma_k$ is fully determined. By solving the two sets of flow equations around the two local minima and comparing the obtained two effective potentials, the true ground state can be fixed.

It should be mentioned that the flow equation for $B_k$ should not be neglected when dealing with the first-order phase transition. According to the definition, the bag constant $B$ is the potential difference between the two local minima. Since the two local minimal potentials are not fixed for a first-order phase transition, one should take into account its renormalization in the potential expansion.

The initial condition for the four flow
equations and the corresponding gap equation at fixed temperature, chemical potential and magnetic field is the four parameters and the classical field at the ultraviolet momentum $\Lambda$,
$a_\Lambda(T,\mu,B),\ b_\Lambda(T,\mu,B),\ c_\Lambda(T,\mu,B),\ B_\Lambda(T,\mu,B)$ and $\sigma_\Lambda(T,\mu,B)$. Considering the fact that the system at high enough
scale is dominated by the dynamics and
not affected remarkably by the temperature, chemical potential and magnetic field, the temperature, chemical potential and magnetic field
dependence of the parameters at the ultraviolet momentum $\Lambda$
can be safely neglected. Therefore, we take the
initial values $a_\Lambda(T,\mu,B)=a_\Lambda(0,0,0),\ b_\Lambda(T,\mu,B)=b_\Lambda(0,0,0),\ c_\Lambda(T,\mu,B)=c_\Lambda(0,0,0),\ B_\Lambda(T,\mu,B)=B_\Lambda(0,0,0)$, and $\sigma_\Lambda(T,\mu,B)=\sigma_\Lambda(0,0,0)$. They are so
chosen to reproduce the proton charge radius, proton magnetic moment and the ratio of axial-vector to vector coupling at fixed $g$ in vacuum at $k=0$. The initial values corresponding to the different groups of physical parameters in Table \ref{para1} are collected in Table \ref{para2} at the ultraviolet scale
$\Lambda=2000$ MeV. For the results shown in the following, we take the first parameter set in Table \ref{para1} with $g=12.416$. For other parameter sets, similar conclusion can be obtained.
%%%%%%%%%%%%%%%%%%%%%%%%%%%%%%%%%%%%%%%%%%%%%%%%%%%%%%%%%%%%%%%%%%%%%%%%%%%%%%%%%%%%%%%%%%%
\begin{table}[ht]
\begin{center}
\caption{The initial parameters in the FRG calculation.}%($\Lambda=2000 \ {\text {MeV}},\ N_c=3,\ N_f=2$)}
\begin{tabular}{ccccc}
set \ &  $a_\Lambda(\text{fm}^{-2})$\   & $-b_\Lambda(\text{fm}^{-1})$  \ & $c_\Lambda$ \ &  $g$ \\
\hline
1 \ & 1749.51 \ & 6400.35 \ & 16549.2 \ & 12.416  \\
\hline
2 \ & 2733.81 \ & 15263.9  \ & 85174.6 \ & 15.333  \\
\hline
3 \ & 1307.08 \ & 3457.19  \ & 5148.48 \ & 10.957  \\
\hline
4 \ & 1610.3 \ & 4950.51 \ & 11692.3 \ &  12.16  \\
\hline
5 \ & 2928.46 \ & 13477.2 \ & 68802.3 \ & 16.379  \\
\hline
6 \ & 1302.34 \ & 3369.6  \ & 4805.52 \ & 10.963  \\
\hline
7 \ & 1614.94 \ & 4854.62  \ & 11117.1 \ & 12.211  \\
\hline
8 \ & 2963.68  \ & 13287.9 \ & 66526.8 \ & 16.537  \\
\hline
\end{tabular}\\ \label{para2}
\end{center}
\end{table}
%%%%%%%%%%%%%%%%%%%%%%%%%%%%%%%%%%%%%%%%%%%%%%%%%%%%%%%%%%%%%%%%%%%%%%%%%%%%%%%%%%%%%%%%%%%%
%%%%%%%%%%%%%%%%%%%%%%%%%%%%%%%%%%%%%%%%%%%%%%%%%%%%%%%%%%%%%%%%%%%%%%%%
\begin{figure}[!htb]
\begin{center}
\includegraphics[width=7.5cm]{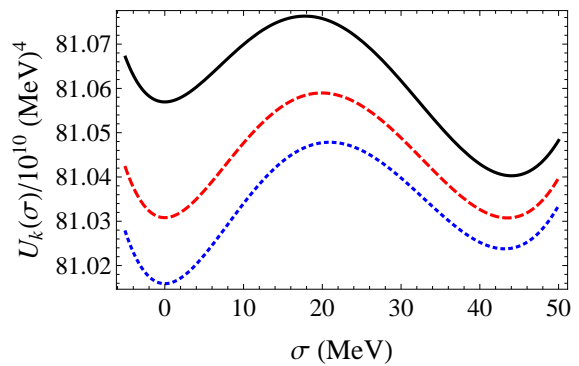}
\caption{The effective potential at different momentum scale in vacuum. The dotted, dashed and solid lines correspond respectively to the scale $k=300,\ 268$ and $0$ MeV. }
\label{fig1}
\end{center}
\end{figure}
%%%%%%%%%%%%%%%%%%%%%%%%%%%%%%%%%%%%%%%%%%%%%%%%%%%%%%%%%%%%%%%%%%%%%%%%

How to choose the value of the ultraviolet scale $\Lambda$ should be carefully discussed in effective models. In principle, the value of $\Lambda$ should be large enough to guarantee the saturation of the obtained physical result at $k=0$. However, in models including hadrons as elementary constituents, the momentum scale can not go beyond the scale of the model itself where the hadrons are well defined. This means that the momentum scale should be restricted in a reasonable region. In the spirit of renormalization group, when more and more fluctuations are involved in the calculation through the momentum scale $k$ approaching from $\Lambda$ to $0$, the phase transition should happen at some critical scale $k_c$. In the Friedberg-Lee model, this is shown in Fig.\ref{fig1}. At $k$=300 MeV, the minimum of the potential is located at $\sigma=\sigma_{per}=0$, the system is in the perturbative vacuum state with deconfined quark mass $m_q=0$. This state maintains until $k\to k_c=268$ MeV. The first-order phase transition happens at $k_c$ with two minima located at $\sigma=\sigma_{per}=0$ and $\sigma=\sigma_{phy}=44$ MeV. For $k<k_c$, $U_{k<k_c}(\sigma_{phy})$ becomes the true minimum and the system is in the physical vacuum state with confined quark mass $m_q=g\sigma_{phy}=547$ MeV. The solid line is the final potential at $k=0$ with maximum fluctuations included in the calculation. From the discussion on the phase transition in vacuum during the evolution of the flow equations, the ultraviolet momentum should be larger than the critical scale $\Lambda>k_c$. Otherwise, there will be no phase transition at finite temperature, density and magnetic field. In the following calculation, we take $\Lambda=2000$ MeV.

\begin{figure}[!htb]
\begin{center}
\includegraphics[width=7.5cm]{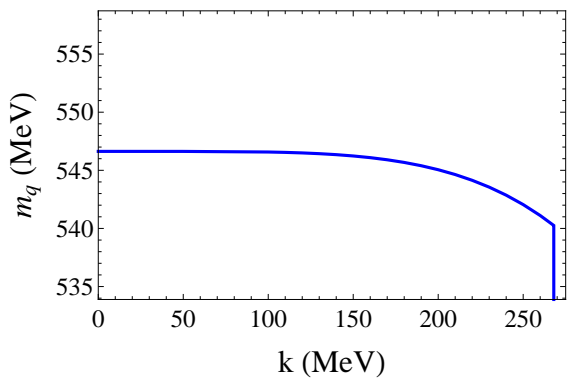}
\caption{The quark mass $m_q$ as a function of the momentum scale $k$ in vacuum. }
\label{fig2}
\end{center}
\end{figure}

Fig.\ref{fig2} shows the evolution of the dynamical quark mass $m_q$ as a function of scale $k$ in vacuum. The quarks are massless before the critical scale $k>k_c$, corresponding to the deconfined state. At the critical scale $k_c$, the mass jumps up suddenly, indicating a first-order phase transition between deconfinement and confinement. Note that, after the phase transition the quark mass is not a constant but changes gradually with the momentum scale $k$. This means that while the perturbative vacuum is always located at $\sigma=0$, the location of the physical vacuum changes smoothly.
%%%%%%%%%%%%%%%%%%%%%%%%%%%%%%%%%%%%%%%%%%%%%%%%%%%%%%%%%%%%%%%%%%%%%%%%
%%%%%%%%%%%%%%%%%%%%%%%%%%%%%%%%%%%%%%%%%%%%%%%%%%%%%%%%%%%%%%%%%%%%%%%%

\begin{figure}[!htb]
\begin{center}
\includegraphics[width=7.5cm]{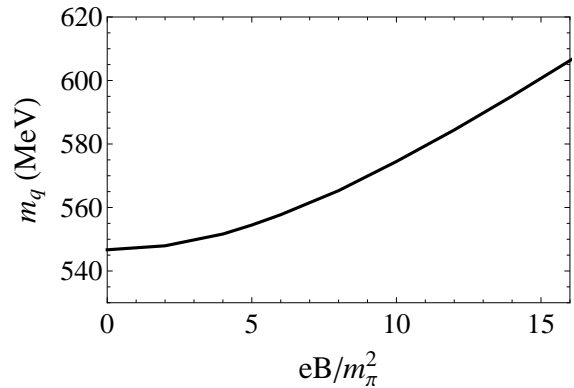}
\caption{The quark mass $m_q$ as a function of magnetic field at zero temperature and baryon density $T=\mu=0$. }
\label{fig3}
\end{center}
\end{figure}

Now we solve the flow equations at finite temperature, density and magnetic field and determine the deconfinement phase transition point. Given fixed temperature, density and magnetic field, the dynamical quark mass is extracted from the evolution of the flow equations from $k=\Lambda$ to $k=0$. Fig.\ref{fig3} shows the quark mass at $k=0$ as a function of the magnetic field at $T=\mu=0$. With increasing magnetic field, the quark mass goes up continuously from $547$ MeV in vacuum to $606$ MeV at $eB=16 m_\pi^2$. The increasing quark mass indicates that, in an external magnetic field quarks are more tightly bound in hadrons and it becomes impossible to break through the confinement by only magnetic field effect.

\begin{figure}[!htb]
\begin{center}
\includegraphics[width=7.5cm]{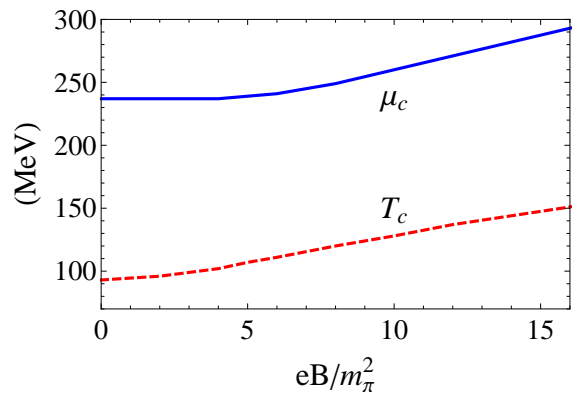}
\caption{The critical temperature $T_c$ at $\mu=0$ and critical chemical potential $\mu_c$ at $T=0$ as functions of magnetic field. }
\label{fig4}
\end{center}
\end{figure}

Fig.\ref{fig4} displays the critical temperature $T_c$ at vanishing baryon density and critical baryon chemical potential $\mu_c$ at vanishing temperature as functions of magnetic field. They both increase monotonously with $B$, clearly indicating magnetic catalysis.
%%%%%%%%%%%%%%%%%%%%%%%%%%%%%%%%%%%%%%%%%%%%%%%%%%%%%%%%%%%%%%%%%%%%%%%%
%%%%%%%%%%%%%%%%%%%%%%%%%%%%%%%%%%%%%%%%%%%%%%%%%%%%%%%%%%%%%%%%%%%%%%%%
%%%%%%%%%%%%%%%%%%%%%%%%%%%%%%%%%%%%%%%%%%%%%%%%%%%%%%%%%%%%%%%%%%%%%%%%
%%%%%%%%%%%%%%%%%%%%%%%%%%%%%%%%%%%%%%%%%%%%%%%%%%%%%%%%%%%%%%%%%%%%%%%%

In summary, the deconfinement phase transition under external magnetic field is investigated by applying the functional renormalization group method to the Friedberg-Lee model. By expanding the effective potential around its two local minima and making comparison of them, which is a generalization of the usually used potential expansion method for continuous phase transitions, we determined the true minimum and the first-order phase transition point. Both the critical temperature and baryon density increase with the magnetic field strength, showing that the deconfinement phase transition in the model becomes more difficult in the external magnetic field. \\

\noindent {\bf Acknowledgement:} I thank Pengfei Zhuang and Hong Mao for useful discussions. The work is supported by the NSFC Grant 11405122 and China Postdoctoral Science Foundation Grant 2014M550483.

\end{document}